\documentclass[pra,aps,twocolumn]{revtex4}

\usepackage{graphicx}
\usepackage{dcolumn}
\usepackage{bm}

\newcommand{\virg}[1]{``#1'' }
\newcommand{\eq}[1]{Eq.~(\ref{#1})}
\newcommand{\fig}[1]{Fig.~\ref{#1}}
\newcommand{\ket}[1]{|#1\rangle}

\begin{document}

\title{Extended Bose Hubbard model of interacting bosonic atoms \\
       in optical lattices: from superfluidity to density waves}
\author{G. Mazzarella}
\author{S. M. Giampaolo}
\author{F. Illuminati}
\affiliation{Dipartimento di Fisica ``E. R. Caianiello'',
Universit\`a di Salerno, Coherentia CNR-INFM, and INFN Sezione di
Napoli, Gruppo Collegato di Salerno, Via S. Allende, 84081
Baronissi (SA), Italy}

\begin{abstract}
For systems of interacting, ultracold spin-zero neutral
bosonic atoms, harmonically trapped and subject to an optical
lattice potential, we derive an Extended Bose Hubbard (EBH) model
by developing a systematic expansion for the Hamiltonian of the
system in powers of the lattice parameters and of a scale
parameter, the {\it lattice attenuation factor}. We identify the
dominant terms that need to be retained in realistic experimental
conditions, up to nearest-neighbor interactions and nearest-neighbor
hoppings conditioned by the on site occupation numbers.
In mean field approximation, we determine
the free energy of the system and study the phase diagram both at
zero and at finite temperature.
At variance with the standard on site Bose Hubbard
model, the zero temperature phase diagram of the EBH model
possesses a dual structure in the Mott insulating regime. Namely,
for specific ranges of the lattice parameters, a density wave
phase characterizes the system at integer fillings, with domains
of alternating mean occupation numbers that are the atomic
counterparts of the domains of staggered magnetizations in an
antiferromagnetic phase. We show as well that in the EBH model, a
zero-temperature quantum phase transition to pair superfluidity is
in principle possible, but completely suppressed at lowest order
in the lattice attenuation factor.
Finally, we determine the possible occurrence of the different
phases as a function of the experimentally controllable lattice
parameters.
\end{abstract}

\date{December 7, 2005}

\maketitle

\section{Introduction}

In the last years we have witnessed a spectacular acceleration in
the experimental manipulation of neutral atoms in optical lattices
\cite{Greiner1,Greiner2,Greiner3,Orzel} that has opened the way to
the simulation of quantum complex systems of condensed matter
physics, such as high-$T_C$ superconductors, Hall systems, and
superfluid $^4He$, thanks to the extreme flexibility and
controllability of such atom-optical systems \cite{Anglin}.
Optical lattices are stable periodic arrays of microscopic
potentials created by the interference patterns of intersecting
laser beams \cite{Jessen1}. Atoms can then be confined in
different lattice sites, and by varying the strength of the
periodic potential it is possible to tune the interatomic
interactions with great precision and to enhance them well into
the regime of strong correlations, even in the dilute limit. The
transition to a strong coupling regime can be realized by
increasing the depth of the lattice potential wells, a quantity
that is directly proportional to the intensity of the laser light
which, in turn, is an experimental parameter that can be
controlled with great accuracy. For this reason, besides the
fundamental interest for the investigation of quantum phase
transitions \cite{Sachdev,van Oosten,Fisher,Sheshadri} and other
collective quantum phenomena
\cite{Ruostekoski,Hofstetter,Paredes,Recati,Buchler,Faber},
optical lattices have become an important tool in applications
ranging from laser cooling \cite{Kerman} to quantum control and
information processing \cite{Jessen2,Duan}, and quantum
computation \cite{Deutsch,Jaksch2,Garcia,Dorner,Pachos}.

The theory of neutral bosonic atoms in optical lattices has been
originally developed, in its simplest framework, by assuming that
the atoms are confined to the lowest Bloch band of the periodic
optical potential \cite{Jaksch1,Zwerger}. It is then
straightforward to show that in this approximation the system is
very well described by the standard, on site Bose Hubbard model.
In such a model a  superfluid-Mott insulator (SF-MI) transition is
predicted to occur when the energy gap between the local ground
state and the first excited levels becomes comparable to the
hopping energy between adjacent lattice sites \cite{Sachdev,van
Oosten,Fisher,Sheshadri}. Moreover, no major qualitative changes
appear when higher excited energy levels of the external trapping
potentials are considered \cite{GIMDS}. This prototypical quantum
phase transition can be realized experimentally, for instance by
manipulating the strength of the lattice potential, which results
in a controlled change of the kinetic (hopping) energy term
\cite{Anglin}. Following this kind of approach, the
superfluid-Mott insulator quantum phase transition has been
realized in a series of beautiful experiments, by loading an
ultracold atomic Bose-Einstein condensate in a three-dimensional
optical lattice \cite{Greiner3}. At finite temperature and in the
strong coupling regime, the quantum phase transition is smeared in
a classical transition from a completely disordered regime to an
ordered, superfluid phase (SF) \cite{Sheshadri,GIMDS}. Finally,
if more complicated geometrical settings such as optical superlattices,
networks, and graphs, are considered, interesting unconventional
interplays between the different quantum phases may appear
\cite{Rothcolour, Penna1, Penna2, Burioni1, Burioni2}.

In the on site Bose Hubbard (BH) model all the terms deriving from
two-body interactions between the bosonic atoms are neglected
except the local ones taking place on a same site of the optical
lattice. This is often an excellent approximation. However,
microscopic interactions are in general of finite range, and it
may be interesting to address the problem of the physical picture
that emerges if one takes into account two-body interactions
between sites at non-vanishing distance and/or hopping amplitudes
beyond nearest-neighbor sites. This problem has been extensively
studied in the case of charged fermions, which are endowed with
very long range Coulomb interactions, usually by introducing
extended Fermi Hubbard models with nearest-neighbor interaction
terms \cite{Efetov,Emery1,Emery2,Robaszkiewicz1,Robaszkiewicz2}.
Extended Fermi Hubbard models are prototypical in theoretical
condensed matter physics, because they are believed to be more
realistic approximations to the true inter-electronic interactions
and moreover exhibit richer phase diagrams when compared to the
standard Hubbard model, especially in low spatial dimensions
\cite{Emery3,Solyom,Lin}.

In the present paper, we extend this kind of analysis to the
bosonic case, and we introduce and study a type of extended Bose
Hubbard model for the description of interacting bosons in regular
lattices. Our study, although general, will be mainly concerned
with the analysis of realistic physical systems such as dilute
ensembles of interacting spin-zero bosonic neutral atoms subject
to an optical lattice potential and spatially confined by a slowly
varying harmonic trapping. Confining our attention to the lowest band
of the optical lattice potential, we construct an effective
extended Bose Hubbard Hamiltonian, that emerges when
we take into account the terms deriving from the boson-boson
interactions both on the same lattice site and on pairs of nearest
neighbors. Our aim is then to analyze the phase transitions that
may occur in such systems both at zero and at finite temperature,
and determine the general features of the associated phase
diagrams.

The paper is organized as follows: In Sec.II we set the general
notations, the needed tools, and derive the model Hamiltonian
from the underlying microscopic dynamics in second quantization. We
identify an exponential lattice scale factor, the {\it lattice
attenuation factor} that gives rise to a natural expansion
parameter for the Hamiltonian of the system. In this way we are
able to determine all the dominant terms that need to be retained
in a power series expansion of the different energy contributions,
that add to the terms of the standard Bose Hubbard (BH) model
Hamiltonian. These additional terms can be easily identified and
ordered in increasing powers of the lattice attenuation factor and
include nearest-neighbor interactions, nearest-neighbor hoppings
of single atoms conditioned by the on site occupation number, and
nearest-neighbor hoppings of atomic pairs (bosonic pairs).
The ensuing Hamiltonian defines a type of extended Bose Hubbard (EBH)
model of interacting bosons in the lowest Bloch band. Other
types of EBH models can be obtained by considering interactions
with far neighbors, and can be related to the fractional quantum
Hall effect \cite{Heiselberg}. Other different types of Hubbard-like
models can also be obtained, at least for one-dimensional fermionic
systems, by implementing appropriate mode expansions of the
second-quantized fermionic field operators, as recently shown
by Massel and Penna \cite{Masselponna}.

The quantum phase diagrams of different types of EBH models have been
extensively investigated by means of quantum Monte Carlo techniques.
In these studies the possible existence of
density wave and supersolid phases in $1$ and in $2$
spatial dimensions has been carefully discussed \cite{Batrouni1,Batrouni2}.
In the present work we set up a formalism that allows to study 
in detail the explicit dependence of EBH models on the physical 
parameters of realistic systems of ultracold neutral atoms in
optical lattice potentials and external confining harmonic 
potentials. 
Equipped with these tools, we can then establish the range 
of values of the Hamiltonian parameters that should be reached
experimentally for the possible observation of the
new quantum phases predicted by EBH models, such as the density
wave Mott insulating regime. Our treatment does not involve 
issues of metastability and lifetime for states defined on excited 
energy bands; it allows the study of the phase diagram both at zero and 
at finite temperature; and, finally, is based on a systematic expansion 
that connects any EBH model to the underlying microscopic many-body
dynamics in second quantization. 
Moreover, it can be in principle extended to more general situations, 
including multi-species bosonic systems and Fermi-Bose mixtures of 
bosonic and fermionic atoms.
Other generalizations of the on site Hubbard model are of course
possible by including Bloch bands of higher order in the description. 
Two such possible EBH models with inter-band hoppings and interactions 
have been recently investigated in connection with proposed schemes 
to generate metastable excited states of cold atoms in optical 
lattices \cite{Scarola,Isacsson}.

In Sec.III, we analyze the phase diagram of the extended Bose
Hubbard model (EBH) in the standard grand canonical formalism. We
first consider the structure of the quantum phases at zero
temperature, when all the kinetic terms vanish (strong coupling
regime). Remarkably, in this case the model is mapped into a
quantum antiferromagnetic Ising model in the presence of an
external magnetic field. As it is well known, this model undergoes
a quantum transition between a ferromagnetic and an
antiferromagnetic phase \cite{Sublattices}. As a consequence, in a
small range of parameters (very strong coupling regime) for a
system of bosons in an optical lattice, there exists a new
insulator phase in which the atomic density is not constant on
each lattice site. This new quantum phase can be seen as a Density
Wave Mott Insulator (DWMI), and is of course absent in the on site
BH model that can sustain only a Pure Mott Insulator (PMI) phase
at constant density. In the DWMI phase, there appear two
alternating domains (sublattices) of different mean on site
occupation numbers, that are the atomic counterparts of the
domains of different staggered magnetization in the
antiferromagnetic phase of the quantum Ising model. Next, we
reintroduce the kinetic terms in the EBH Hamiltonian by resorting
to Bogoliubov-like and mean field approximations. In this
framework three real-valued order parameters emerge: the
conventional single boson (single atom) superfluid order
parameter, a new bosonic pair superfluid order parameter, and
finally the mean number of bosons (atoms) per site. This
parameters are not all independent from each other, due to the
existence of  physical constraints on thermodynamic stability and
on the PMI-DWMI phase separation.

In Sec.IV, we present qualitative and analytical studies of the
different possible phases in the presence of non vanishing kinetic
energy terms. In this case, minimization of the free energy with
respect to the different order parameters yields that the only
superfluid transition which can occur is the one ruled by the
single atom SF order parameter. We analyze the behavior or the
SF-PMI and the SF-DWMI quantum phase transitions at zero
temperature, and the behavior of the SF single boson order
parameter at finite temperature. We study the critical temperature
of the disordered-SF phase transition as a function of the filling
factor and of the lattice depth. Then, starting from finite
temperatures, we analyze the possibility to recover the SF-PMI and
the SF-DWMI quantum phase transitions by determining the behavior
of the critical energy gap in the limit of vanishing temperature,
as a function of the lattice depth and for different values of the
trapping frequencies. By comparing the phase diagrams obtained
for different values of the Hamiltonian parameters, we discuss 
the possibility of observing experimentally the zero-temperature
transition to the SF-DWMI phase in systems of neutral atoms
loaded in optical lattice potentials. 
Finally, in Sec.V we give some concluding comments on the obtained 
results and discuss some possible directions for future research.

\section{General Setting}

The microscopic Hamiltonian for an ensemble of bosonic atoms that
are confined by a slowly varying external harmonic trapping
potential and subject to an additional optical lattice can be
written as
\begin{equation}
 \label{fullhamiltonian}
 \hat{H} \, = \, \hat{T} \, + \, \hat{V} \, + \, \hat{W} \; ,
\end{equation}
where $\hat{T}$ is the kinetic energy term that reads
\begin{equation}
 \label{Hamiltonian operator T}
  \hat{T} \, = \, -
 \frac{\hbar^{2}}{2 m} \int d \vec{r} \, \hat{\Psi}^{\dagger}(\vec{r})
 \nabla^{2} \hat{\Psi}(\vec{r}) \; ,
\end{equation}
with $\hat{\Psi}(\vec{r})$ being the bosonic annihilation field
operator at point $\vec{r}$. On the other hand $\hat{V}$
represents the contribution of the external potential to the
energy:
\begin{equation}
 \label{Hamiltonian operators VF}
\hat{V} \, = \, \int d\vec{r} \, \hat{\Psi}^{\dagger}(\vec{r})
\left(V_{H}(\vec{r}) \, + \, V_{opt}(\vec{r}) \right)
\hat{\Psi}(\vec{r}) \; .
\end{equation}
In concrete situations, spatial confinement of the atoms is
provided by the quadrupolar anisotropic trapping magnetic field
that leads to an harmonic trapping potential of the form
\begin{equation}
 \label{harmonic potential}
 V_H \, = \, \frac{m\omega^{2}}{2}\left(x^{2}+\lambda^{2}y^{2}+\lambda^{2}z^{2}\right) \; ,
\end{equation}
where $\omega$ is the frequency associated to the harmonic trap in
the $x$ direction and $\lambda$ is the anisotropic coefficient,
i.e. the ratio between the frequency in the $yz$ plane and the
frequency in the $x$ direction. The most common experimental
settings are realized in the so-called \mbox{\virg{cigar-shaped}}
configuration $(\lambda \gg 1)$. Correspondingly, the second
contribution to the potential energy is a \mbox{$1$-D} periodic
potential needed to realize the optical lattice along the axis of
the ``cigar'':
\begin{equation}\label{Optical potential}
  V_{opt}(x) \, = \, V_{0}\sin^{2}\left( \frac{\pi x}{a}  \right) \; ,
\end{equation}
where $V_{0}$ is the maximum amplitude of the light shift
associated to the intensity of the laser beam and $a$ is the
lattice spacing related to the wave vector $k$ of the standing
laser light by $k=\pi/a$.

Finally, the third contribution is the local, contact two-body
interaction
\begin{equation}
\label{twobodyinteraction} \hat{W} \, = \, \frac{g_{BB}}{2}\int
d\vec{r} \hat{\Psi}^{\dagger}(\vec{r})
\hat{\Psi}^{\dagger}(\vec{r})\hat{\Psi}(\vec{r})
\hat{\Psi}(\vec{r}) \; ,
\end{equation}
in which the interaction coupling $g_{BB}=4\pi\hbar^{2}a_{BB}/m$
where $m$ is the atomic mass and $a_{BB}$ is the atom-atom
(boson-boson) $s$-wave scattering length. In the following, we
will always assume boson-boson repulsion, i.e. $a_{BB}>0$.

In the presence of a strong optical lattice and a sufficiently
shallow external confinement in the $x$ direction, so that at any
lattice site its value can be considered constant, the bosonic
field operators can be expanded in the basis of the
single-particle Wannier wave-functions localized at each lattice
site $x_i$. Since the typical interaction energies involved are
normally not strong enough in order to excite higher vibrational
states, we can retain only the the lowest vibrational state in
each lattice potential well (single-band approximation). In the
case of stronger external confinements, or interactions, one
should include higher Bloch bands as well in the expansion of
field operators, a case we do not consider in the present context.
Moreover, as far as the harmonic trapping potential is concerned,
we have shown in a previous work \cite{GIMDS} that the
introduction of higher energy levels of the harmonic oscillator
does not modify the basic phenomenology of the system. Under these
conditions, we can avoid working directly with the exact Wannier
wave functions and replace them, with an excellent degree of
fidelity, with their harmonic-oscillator approximations at each
optical lattice well. Then, the Wannier wave functions
$w(\vec{r})$ factorize in the product of harmonic oscillator
states in each direction:
\begin{equation}
\label{wannierexpansion} \hat{\Psi}(\vec{r}) \, = \, \sum_{i}
\hat{a}_{i} w(x-x_{i}) w(y) w(z) \; ,
\end{equation}
where $x_i$ is the center of the $i$-th lattice well and
$\hat{a}_{i}$ is the bosonic annihilation operator acting at the
\mbox{$i$-th} lattice site. In each lattice potential well, the
Wannier local ground states are Gaussians in the harmonic
approximation:
\begin{eqnarray} \label{Harmonic oscillator eigenstate}
 w(x-x_{i})&=&\frac{1}{\sqrt{l_{x} \sqrt{\pi}}}
 \exp\left[\frac{-(x-x_{i})^{2}}{2l^{2}_{x}}\right] \nonumber \; , \\
 w(y)&=&\frac{1}{\sqrt{L_{\perp} \sqrt{\pi}}}
 \exp\left[\frac{-y^{2}}{2L^{2}_{\perp}}\right] \; ,\\
 w(z)&=&\frac{1}{\sqrt{L_{\perp} \sqrt{\pi}}}
 \exp\left[\frac{-z^{2}}{2L^{2}_{\perp}}\right] \; . \nonumber
\end{eqnarray}
In \eq{Harmonic oscillator eigenstate} we have introduced the
harmonic oscillator lengths of the ground state in the $y$ and $z$
directions $L_{\perp} = \sqrt{\hbar/(m\lambda \omega)}$, and the
oscillator length in the harmonic approximations of the periodic
potential $l_{x}$ that reads, as a function of the lattice
parameters, \mbox{$l_x = \left( a^4 E_R / (\pi^4 V_0 )
\right)^{1/4}$}, where $E_R = (\pi \hbar)^2/2 a^{2} m$ is the
lattice recoil energy. In this paper we will consider the physical
situation of very shallow trapping potentials, such that $L_{x}
\equiv L_{\perp}\sqrt{\lambda} \gg a M$, with $M$ denoting the
total number of lattice sites. As a consequence, the local density
approximation (LDA) can be applied. Therefore, when exploiting the
expansion \eq{wannierexpansion} to map the full microscopic Hamiltonian
\eq{fullhamiltonian} into its lattice version, we will discard all
terms that are of order $(aM/L_{x})^2$ or higher. Qualitatively,
this means neglecting those nonlocal effects that are induced by the
presence of the trapping potentials, such as site-dependent hopping terms.
The latter can become important in regions of the lattice very far out
of the central core of the harmonic trap. However, the typical
experimental situations involve only that part of the lattice that lies
well inside the central core of the slowly varying confining
potential \cite{Ferlaino}. We can then write down
the translationally invariant lattice
version of Hamiltonian \eq{fullhamiltonian} in the form
\begin{equation}
\label{latticeh1}
\hat{H} \, = \, -\frac{1}{2}\sum_{i,j} t _{i,j}\hat{a}^{\dagger}_{i}
\hat{a}_{j} \, + \, \frac{1}{2}\sum_{i,j,k,l} U_{i,j,k,l}
\hat{a}^{\dagger}_{i}\hat{a}^{\dagger}_{j}\hat{a}_{k}\hat{a}_{l} \; .
\end{equation}
In \eq{latticeh1} $U_{i,j,k,l}$ is the two-body interaction
strength that involves four sites of the lattice that depends on
the relative distance between the sites involved. Recalling the
expression of the two-body interaction \eq{twobodyinteraction},
together with the form of the bosonic field operator
\eq{wannierexpansion} and of the lattice wave functions
\eq{Harmonic oscillator eigenstate}, one has
\begin{equation}
\label{U function d} U_{i,j,k,l} \, = \, U_{0}
\varepsilon^{\gamma/2} \; ,
\end{equation}
where
\begin{equation}
U_{0} \, = \, (2\pi)^{-\frac{3}{2}}
\frac{g_{BB}}{l_{x}L_{\perp}^{2}} \; , \label{uconzero}
\end{equation}
is the local interaction strength, i.e. the amplitude of the
interaction when $i=j=k=l$,
\begin{equation}
\varepsilon \, = \, \exp\left(-a^{2}/4l_{x}^2\right) \; ,
\label{epsilon}
\end{equation}
is the {\em lattice attenuation factor}, and
\begin{eqnarray}
\gamma \, & = & \, (i-j)^2 \, + \, (i-k)^2 \, + \, (i-l)^2 \nonumber \\
& + & (j-k)^2 \, + \, (j-l)^2\, + \, (k-l)^2 \; ,
\label{fourdistance}
\end{eqnarray}
is the \virg{four-site distance} relative to all possible
independent pairs of sites that can be chosen out of a set of four
sites. It is worth noticing that $\varepsilon$ can be re-expressed
in the form $\varepsilon=\exp(-\pi^2\sqrt{s}/4)$, i.e. in terms of
an experimentally measurable and tunable quantity, the depth $s$
of the lattice wells: $s=V_0/E_R$.

On the other hand, in \eq{latticeh1}, $t_{i,j}$ is the strength of
the contributions of the kinetic and the external potential terms
to the energy. For $i=j$ it gives rise to a constant zero point
energy term that can be discarded by redefining the zero of the
energy; if $i \neq j$ it represents the probability amplitude for
an atom to tunnel from the \mbox{$i$-th} lattice site to the
\mbox{$j$-th one} along the $x$ direction. Obviously, also this
probability amplitude is a function of the distance between the
involved sites, but it is impossible to write for it a closed
analytical formula like \eq{U function d}. However, from the form
of \eq{Harmonic oscillator eigenstate} it is easy to show that $t
_{i,j}$ is still proportional to some positive power of the
lattice attenuation factor $\varepsilon$, with the exponent
depending only on the distance between the sites. Hence, in
general one can write
\begin{equation}\label{J function d}
\label{J} t _{i,j} \, = \, J_{|i-j|}\varepsilon^{(i-j)^2} \; ,
\end{equation}
where $J_{|i-j|}$ decreases as a polynomial function of the
modulus of the distance between the sites. Taking into account the
form of the kinetic energy \eq{Hamiltonian operator T}, the forms
of the external potentials \eq{Hamiltonian operators VF},
\eq{harmonic potential}, and \eq{Optical potential}, together with
the expression of the bosonic field operators
\eq{wannierexpansion} and of the lattice wave functions
\eq{Harmonic oscillator eigenstate}, for $|i-j|=1$ (nearest
neighbors) one has
\begin{widetext}
\begin{equation}\label{J1}
 J_1 \; = \; V_0\left(\frac{\pi^2}{2}-1-e^{-(\pi^{2} l_x^2)/a^2}\right)
-\frac{m \omega^2 l_x^2}{2} -2 \lambda \omega \hbar
 -\frac{\hbar^2}{2 l_x^2 m} \; \simeq \; 2 V_0
 \left(\frac{\pi^2}{4}-1\right) \; .
\end{equation}
\end{widetext}
Writing the single particle hopping amplitude as in \eq{J function
d} remarks the fact that the one-body contributions to the energy
are function of the lattice attenuation factor $\varepsilon$ as
well. In typical experimental situations, the lattice spacing $a$
is usually much larger than the local ground state length $l_{x}$
at each lattice site. Hence $\varepsilon \ll 1$, and this fact
allows to exploit the lattice attenuation factor as a meaningful
dimensionless expansion parameter. The first two nontrivial
contributions to the one-body part of the energy are proportional
to $\varepsilon$, corresponding to $|i-j|=1$ and $\varepsilon^4$,
corresponding to $|i-j|=2$. The first contribution is the usual
nearest-neighbor hopping of the standard Bose-Hubbard model, while
the second one is a next-to-nearest-neighbor hopping term.

Concerning the two-body contributions to the energy, the
classification in powers of $\varepsilon$ looks in principle more
complicated. However, it is easy to see that all the terms related
to pairs of nearest-neighbor sites are proportional to powers
$\varepsilon^{l}$ at most of order $l=2$. On the other hand,
energy terms involving pairs of next-to-nearest-neighbors or
triples of three adjacent sites are always smaller, because the
leading terms of this two classes of energy contributions are,
respectively, proportional to $\varepsilon^{6}$ and
$\varepsilon^3$. Hence, in the lattice Hamiltonian description of
interacting bosonic atoms in periodic optical potentials, we need
to consider, in first approximation, only the energy terms
proportional to $\varepsilon^{l}$ with $l \leq 2$. At this order
of approximation, the lattice Hamiltonian \eq{latticeh1} reads,
with the terms ordered in increasing powers of $\varepsilon$,
\begin{widetext}
\begin{eqnarray}
\label{Hamiltonian} && \hat{H} \, = \, \frac{U_{0}}{2}
\sum_{i}\hat{n}_i(\hat{n}_i-1) \, - \,
\frac{J_1}{2}\varepsilon\sum_{i}(\hat{a}^{\dagger}_{i}\hat{a}_{i+1}
+ H. c.) \, + \, U_{0} \varepsilon^{\frac{3}{2}} \sum_{i}
\left[(\hat{a}^{\dagger}_{i} \hat{n}_i \hat{a}_{i+1}
+ \hat{a}^{\dagger}_{i-1} \hat{n}_{i} \hat{a}_{i}) + H. c.\right] \nonumber \\
&& \nonumber \\
&& + \, 2 U_{0} \varepsilon^2\sum_{i}\hat{n}_{i}\hat{n}_{i+1} \, +
\, {U_{0}} \varepsilon^2\sum_{i}\left(
\hat{A}^{\dagger}_{i}\hat{A}_{i+1}+H.c. \right) \; ,
\end{eqnarray}
\end{widetext}
where  we have introduced both the on site occupation number
operator $\hat{n}_i=\hat{a}_{i}^{\dagger}\hat{a}_{i}$ and the on
site pair annihilation operator $\hat{A}_{i}\equiv
\hat{a}_{i}^{2}$. The first two terms of \eq{Hamiltonian}
represent the usual BH Hamiltonian with on site interaction and
single-atom nearest-neighbor hopping. The remaining terms give the
corrections to this model, associated to higher powers of the
lattice attenuation factor $\varepsilon$. The term proportional to
$\varepsilon^{3/2}$ is the single-boson nearest-neighbor hopping
conditioned by the on site occupation; the first term proportional
to $\varepsilon^{2}$ is the density-density nearest-neighbor
interaction; and, finally, the second term proportional to
$\varepsilon^{2}$ is the nearest-neighbor hopping of pairs of
bosons.

\section{The free energy}

When dealing with systems of interacting bosons, it is convenient
to work in the framework of the grand canonical ensemble
\cite{Fisher,Sheshadri,van Oosten,GIMDS}. Let us then introduce
the grand canonical Hamiltonian $\hat{K}$
\begin{equation}
\label{grancanonicaloperator} \hat{K} \, = \, \hat{H} \, - \, \mu
\sum_i{\hat{n}_i} \; ,
\end{equation}
where $\mu$ is the chemical potential needed to fix the average
total number of bosons in the lattice. All the summations entering
in \eq{grancanonicaloperator} can be arranged in two different
sets $(\hat{K}_l)$ and $(\hat{K}_{int})$. The first one,
$(\hat{K}_l)$, contains all ``local'' terms that depend only on
the on site occupation number operators $\hat{n}_i$ and
$\hat{n}_{i+1}$. The second one, $(\hat{K}_{int})$ contains all
the ``non local'' hopping terms. According to this grouping, one
can write
\begin{equation}
\label{grancanonicaloperator1} \hat{K} \, = \, \hat{K}_l \, + \,
\hat{K}_{nl} \; .
\end{equation}
We will first analyze the EBH model when the second, the third
and the last terms of the right-hand side of \eq{Hamiltonian},
i.e. the kinetic contributions to energy, may be neglected and
only the local terms are retained (in a sense that will be
clarified below).

\subsection{Local energy terms and mapping to a quantum Ising antiferromagnet}

Considering \eq{Hamiltonian} and \eq{grancanonicaloperator}, the
local energy part in \eq{grancanonicaloperator1} reads
\begin{equation}
\label{localgrancanonicalterm} \hat{K}_l =  \frac{U_{0}}{2}
\sum_{i}\hat{n}_i(\hat{n}_i-1)-\mu \sum_i{\hat{n}_i}+2 U_{0}
\varepsilon^2\sum_{i}\hat{n}_{i}\hat{n}_{i+1} \; .
\end{equation}
Since in the \virg{local} part of the Hamiltonian we include the
nearest-neighbor interaction term, we should qualify that here, by
``local'' we mean all effects that do not involve particle
exchange between sites.

To fix techniques and notations, we first briefly recall how to
determine the energy spectrum of the local part of the Hamiltonian
in the standard BH model, i.e. when we neglect the term
proportional to $\varepsilon^2$ in \eq{localgrancanonicalterm}. In
this case, it is well known that each term in the sum of on site
interaction energies reaches its minimum for \mbox{$n_i=n^*$} with
\begin{equation}
\label{min} n^* \, = \, \frac{1}{2} \, + \, \frac{\mu}{U_{0}} \; .
\end{equation}
This trivial observation naturally leads to introduce the complete
orthonormal Fock basis $\{\ket{n_i}\}$ and determines an obvious
but important classification. If $n^*$ is close to an integer
value, say $n_0$, we have that the energy gap $\Delta$ between the
ground state $\ket{n_0}$ and the first excited state is of the
order of the coupling constant $U_0$. This, incidentally,
justifies neglecting, in first approximation, any correction
proportional to any power $l\geq 2$ of $\varepsilon$. In this
situation, at zero temperature, the system is in a Mott-Insulator
phase with exactly $n_0$ atoms per site. On the other hand, if
$n^*$ is closer to a half-integer value, then the energy gap
$\Delta$ can be comparable with other contributions to the
interaction energy, and one can write
\begin{eqnarray}
\label{minbis} n^* \, = \, \frac{1}{2} \, + \, n_0 \, + \,
\frac{2\Delta}{U_{0}} \; ,
\end{eqnarray}
where $n_0$ is integer, $|\Delta|/U_{0}\ll 1$ and the two number
states nearly degenerate in energy are $\ket{n_0}$ and
$\ket{n_0+1}$. This near degeneracy occurs in pairs: the states
$\ket{n_0+2}$ and $\ket{n_0-1}$ (when it exists, i.e. when $n_0
\geq 1$) are as well nearly degenerate and are separated from
the pair $\{ \ket{n_0}$, $\ket{n_0+1} \}$ by a gap of the order
of $U_0$.
The two nearly-degenerated ground states $\ket{n_0}$
and $\ket{n_0+1}$ are separated from each other
by an energetic distance equal to $2\,|\Delta|$, while the
gap between the two nearly-degenerate first excited
states $\ket{n_0+2}$ and $\ket{n_0-1}$
is equal to $6\,|\Delta|$. These results hold
analogously for the pairs $\{ \ket{n_0+3}$, $\ket{n_0-2} \}$ and
$\{ \ket{n_0+4}$, $\ket{n_0-3} \}$, and so on, as long as the second
element $\ket{n_0 - k}$ of each pair exists. From this
classification, it emerges the fundamental role played by the two
nearly degenerate states $\ket{n_0}$ and $\ket{n_0+1}$. Moreover,
the previous analysis allows to recast the local Hamiltonian
\eq{localgrancanonicalterm} in a very useful form. Introducing the
operator $\hat{m}_i \, \equiv \, \hat{n}_i-(n_0+\frac{1}{2})$ and
fixing the zero of the energy, the local part of the grand
canonical Hamiltonian reads
\begin{equation}
\label{localterm1} \hat{K}_l = \frac{U_{0}}{2}
\sum_{i}\left(\hat{m}_i^{2}-\frac{1}{4}\right) -2\delta
\sum_i{\hat{m}_i}+ K \sum_{i}\hat{m}_{i}\hat{m}_{i+1} \; ,
\end{equation}
where $\delta \equiv  \Delta-2U_{0}\epsilon^2 (n_0+\frac{1}{2})$,
and $K \equiv 2 U_{0}\epsilon^2$.

The above discussion and \eq{localterm1} allow a clear
understanding of the local part of the EBH model at zero
temperature, showing that all the states different from the two
quasi degenerate ground states $\ket{n_0}$ and $\ket{n_0+1}$ do
not contribute. Then, the first term in \eq{localterm1} can be
neglected, and the remaining part of the local Hamiltonian
(\ref{localterm1}) describes an assembly of interacting two-level
systems. Actually, because $K>0$, it is mapped exactly in a spin
$1/2$ antiferromagnetic quantum Ising model in the presence of an
external field $-2\delta$:
\begin{equation}
\label{Ising} \hat{H}_{eq} \, = \,
-2\delta\sum_{i}\hat{\sigma_{i}^{z}} \, + \,
K\sum_{i}\hat{\sigma_{i}^{z}}\hat{\sigma_{i+1}^{z}} \; .
\end{equation}
At zero temperature, this model describes a system that undergoes
a quantum phase transition at the critical field $\delta_c=K/2$.
The ferromagnetic phase $|\delta|>\delta_c$ corresponds to the
Mott-Insulator phase with the same, constant number of atoms $n_0$
on each lattice site. The antiferromagnetic phase
$|\delta|<\delta_c$ corresponds to a density wave phase with $n_0$
atoms on a site and $n_0+1$ on its neighbor. Analogously to the
spin system in the antiferromagnetic phase, the optical lattice
for the bosonic atoms in the density wave phase is divided in two
sublattices of ``staggered'' atomic densities $n_0$ and $n_0+1$.
In the following, we will denote the two phases, respectively by
PMI (Pure Mott-Insulator) and by DWMI (Density Wave
Mott-Insulator).

\subsection{Nonlocal energy terms, ferromagnetic- and antiferromagnetic-like
models, and the mean field free energy}

When the nonlocal hopping terms are reintroduced, the tensor
product states of the local occupation number states (local Fock
states) are no longer eigenstates of the total Hamiltonian. The
true eigenstates cannot be determined analytically, and consistent
approximations must be envisaged to approach the problem in the
product basis of the local states. Let us first rewrite the
nonlocal terms appearing in the grand canonical Hamiltonian
(\ref{grancanonicaloperator1}) in terms of the on site
``magnetization'' operators  $\hat{m}_i$:
\begin{widetext}
\begin{equation}
\label{grancanonicalnonlocale} \hat{K}_{nl} =  -  \frac{J}{2}
\sum_{i}\left(\hat{a}^{\dagger}_{i}\hat{a}_{i+1}+H.c.\right)+U_{0}
\varepsilon^{\frac{3}{2}} \sum_{i}\left( \hat{a}^{\dagger}_{i}
(\hat{m}_i+\hat{m}_{i+1}) \hat{a}_{i+1} +H.c.\right) +\frac{K}{2}
\sum_{i}\left(\hat{A}^{\dagger}_{i}\hat{A}_{i+1} + H.c.\right)\;,
\end{equation}
\end{widetext}
where the ``dressed'' hopping amplitude $J$ reads
\begin{equation}
J \, \equiv \, \varepsilon \left[ J_1 - 2U_{0}
\varepsilon^{\frac{1}{2}} \left( n_0+\frac{1}{2} \right) \right]
\; . \label{hoppingdressed}
\end{equation}
From Eq. (\ref{hoppingdressed}) we see that the density-dependent
part of the hopping amplitude gives a negative contribution if the
boson-boson interactions are repulsive ($U_{0} > 0$). Stability of
the ground state energy thus requires
\begin{equation}
J_1 \, > \, 2U_{0} \varepsilon^{\frac{1}{2}}\left( n_0+\frac{1}{2}
\right) \; . \label{stability}
\end{equation}
Typically, such a stability requirement can be easily
satisfied in most experimental situations, unless one goes to very
large occupation numbers $n_0$ and very strong interaction couplings $U_0$.
Hence, in the following discussions and examples, we will always
consider situations in which the stability condition
\eq{stability} is satisfied.

Before introducing mean field approximations, we first need to
deal with the second term in \eq{grancanonicalnonlocale}, the
conditioned hopping term. This can be done, in a Bogoliubov-like
framework, by replacing the operator $\hat{m}_i$ with its average
value $\chi_i=<\hat{m}_i>$, thus neglecting quantum fluctuations.
This is justified in the situation in which the magnitude of the
on-site interaction amplitude $U_0$ is sufficiently large that the
probability of finding on site occupation numbers that do not fall
in the ranges identified by the pairs $\{n_0$, $n_0+1\}$ and
$\{n_0 +2$, $n_0 -1\}$ is negligible. This is the physical
situation that one usually meets in realistic experimental
conditions. In implementing this approximation we must thus
distinguish between two different instances, according to the
previous discussion on the local terms.

{\it I} -  If one has $|\delta|>K/2$, then the approximate model
describing the system is ferromagnetic-like and, in the limit of
vanishing nonlocal hopping terms, the associated ground state
reduces exactly to the PMI phase. In the ferromagnetic-like model,
the expectation value of the occupation number $\chi_i$ is
constant on all sites of the optical lattice: $\chi_i=\chi \;
\forall i$.

{\it II} - If one has $|\delta|<K/2$, then the approximate model
describing the system is antiferromagnetic-like and, in the limit
of vanishing nonlocal hopping terms, the associated ground state
reduces exactly to the DWMI phase. In the antiferromagnetic-like
model, the expectation value of the on site occupation number
takes opposite values on neighboring sites of the optical lattice:
$\chi_i=\chi, \; \chi_{i+1}=-\chi$ and henceforth the conditioned
hopping term in \eq{grancanonicalnonlocale} vanishes.

Then, in compact notation, the grand canonical Hamiltonian
$\hat{K}^{F}$ for the ferromagnetic-like case and the grand
canonical Hamiltonian $\hat{K}^{A}$ for the antiferromagnetic-like
case read:
\begin{widetext}
\begin{equation}
\label{grancanonico1} \hat{K}^{F,A} =\frac{U_{0}}{2}
\sum_{i}\left(\hat{m}_i^{2}-\frac{1}{4}\right) -2\delta
\sum_i{\hat{m}_i}+ K \sum_{i}\hat{m}_{i}\hat{m}_{i+1} -
\frac{J^{F,A}}{2}
\sum_{i}\left(\hat{a}^{\dagger}_{i}\hat{a}_{i+1}+H.c.\right) +
\frac{K}{2} \sum_{i}\left(\hat{A}^{\dagger}_{i}\hat{A}_{i+1} +
H.c.\right).
\end{equation}
\end{widetext}
Here, $J^{F,A}$ are the single particle hopping amplitude of the
two models. When $|\delta|>\delta_c$, one must take the choice
$J^{F}= J-4 U_{0}\varepsilon^{\frac{3}{2}} \chi$. When
$|\delta|<\delta_c$, one must take the choice $J^{A}= J$.

Before proceeding, we would like to make the following side
observation. The grand canonical operator \eq{grancanonico1} takes
into account both the local and the nonlocal parts of energy. We
have seen that when we can neglect the nonlocal terms (i.e. in the
absence of the kinetic terms), the local part of the Hamiltonian
\eq{localgrancanonicalterm} is mapped in a spin-$1/2$ quantum
Ising model \eq{Ising}. We can exploit a similar mapping also for
the total Hamiltonian \eq{grancanonico1} in the particular limit
when the on site interactions are strong enough that the only
states that contribute are those with $n_{0}$ and $n_{0}+1$ bosons
per lattice site. In this special limit, the nearest neighbor
atomic pair hopping operator $\hat{A}_i^{\dagger}\hat{A}_{i+1}$
does not produce any effect. Hence, by the same identifications
$\hat{a}^{\dagger}_{i}=\hat{\sigma}^{\dagger}_{i}=\hat{\sigma}^{x}_{i}+i
\hat{\sigma}^{y}_{i}$ and
$\hat{m}^{\dagger}_{i}=\hat{\sigma}^{z}_{i}$ that map Hamiltonian
$\hat{K}_{l}$ \eq{localgrancanonicalterm} to the quantum Ising
model \eq{Ising}, the two Hamiltonians $\hat{K}^{F,A}$
\eq{grancanonico1} are mapped in the two spin-$\frac{1}{2}$ $XXZ$
Hamiltonians
\begin{eqnarray}
\label{XXZ} \hat{H}_{XXZ}^{F,A} & = &
-2\delta\sum_{i}\hat{\sigma_{i}^{z}}+K\sum_{i}\hat{\sigma_{i}^{z}}\hat{\sigma_{i+1}^{z}}
\nonumber\\
&& - J^{F,A}\sum_{i}(\hat{\sigma}_{i}^{x}\hat{\sigma}_{i+1}^{x}
+\hat{\sigma}_{i}^{y}\hat{\sigma}_{i+1}^{y}) \; .
\end{eqnarray}
The above limiting mapping of Bose-Hubbard models in $XXZ$
Hamiltonians in external field has been investigated extensively
by van Otterlo {\it et al.} \cite{vanOtterlo}, who predicted the
existence of a supersolid phase in two dimensions.

We are now in a position that allows to introduce mean field
approximations on the generic terms containing pairs of operators
an adjacent sites, namely: $\hat{a}_i^{\dagger}\hat{a}_{i+1}$;
$\hat{A}_i^{\dagger}\hat{A}_{i+1}$; and $\hat{m}_i \hat{m}_{i+1}$.
For the latter term, we must make a bookkeeping for the two
different model Hamiltonians corresponding to
$|\delta|>|\delta_c|$ and $|\delta|<|\delta_c|$. In the first case
we must consider $\chi=<\hat{m}_i> \; \forall \,i$. In the second
case the order parameter has opposite signs on adjacent sites. For
the first two pairs of terms, in order to implement correctly the
mean field approximation, we must make sure that concavity of the
energy holds, guaranteeing that the extremal conditions correspond
to a true minimum of the energy and not to a maximum. For the
single particle hopping, recalling that $-J^{F,A}<0$ we must
consider a uniform order parameter $<a_i>=<a_i^{\dagger}>=\phi \;
\forall \, i$. For the pair hopping term we have that its
amplitude $K$ is always positive; hence, in order to obtain the
right concavity, we need to choose an order parameter that takes
opposite signs on adjacent sites: $<A_i>=\psi, \;
<A_{i+1}>=-\psi$. We have chosen to restrict to real order
parameters even if, due to the non-hermiticity of the involved
operators, in principle complex order parameters would be allowed.
Obviously, the imaginary parts of the order parameters may be
easily taken into account. However, we have verified that even in
these cases the extremal conditions always lead to real results.
For this reason we can restrict our analysis right from the start
to real order parameters, a situation that is in complete analogy
with the one encountered in the study of the standard on site BH
model \cite{Sheshadri,van Oosten}

We can now write down the mean field expressions for the
ferromagnetic-like and antiferromagnetic-like grand canonical
total Hamiltonians (with $M$ being the total number of lattice
sites):
\begin{widetext}
\begin{eqnarray}
\label{grancanonicomfa} \hat{K}^{F,A} &=&\frac{U_{0}}{2}
\sum_{i}\left(\hat{m}_i^{2}-\frac{1}{4}\right) \, - \, J^{F,A}
\sum_{i}\left(\hat{a}^{\dagger}_{i}+\hat{a}_{i}\right) \phi \, -
\, (\delta-K \chi) \sum_{i \in S1} {\hat{m}_i} \, - \,
(\delta \mp K \chi) \sum_{i \in S2} {\hat{m}_i} \nonumber \\
&& \nonumber \\
&+&\frac{K}{2} \sum_{i\in S1}
\left(\hat{A}^{\dagger}_{i}+\hat{A}_{i}\right)\psi \, - \,
\frac{K}{2} \sum_{i \in S2} \left(\hat{A}^{\dagger}_{i}
+\hat{A}_{i}\right)\psi  \, + \, \left(K \psi^2 + J^{F,A}\phi^2
\mp  K \chi^2\right)M \; ,
\end{eqnarray}
\end{widetext}
where the minus sign holds for the ferromagnetic and the plus sign
for the antiferromagnetic case. By $S1$ and $S2$ we denote the two
different sublattices in which the original lattice is split
with regard to the $\psi$ and $\chi$ order parameters (for the
latter, only in the case $2|\delta|<K$).

The grand canonical total Hamiltonians are written down as sums of
local on site energy terms, and the order parameters
$\{\chi,\phi,\psi\}$ must be evaluated self-consistently. In
principle, the Fock spaces associated to the on site occupation
numbers are infinite-dimensional. However, the leading term in
\eq{grancanonicomfa} is the one proportional to $U_0$, so that all
number states with eigenvalue greater than $U_0$ can be neglected,
leading to consider only the set of the four lowest lying states
that include the two nearly degenerate local states $|n_0>$ and
$|n_0+1>$, and the two nearly degenerate local states $|n_0+2>$
and $|n_0-1>$ (when the latter exists, i.e. when $n_0 \geq 1$).

Starting from the two grand canonical Hamiltonians
\eq{grancanonicomfa} we can evaluate analytically the two free
energies $F$ of the system at any inverse of the temperature
$\beta = (k_{B}T)^{-1}$ either in the ferromagnetic-like or in the
antiferromagnetic-like case. One elegant technique to do so is to
resort to the resolvent approach, as illustrated in the Appendix.
Considering the free energy per site $f^{F,A}$ in the
thermodynamic limit, one has:
\begin{widetext}
\begin{eqnarray}
\label{FreeEnergy} f^{F,A} & = & J^{F,A}\phi^{2} \, + \, K\psi^{2}
\, \mp \, K\chi^2 \, - \, \frac{(J^{F,A} \phi)^{2} (n_{0}+1)
+K^{2}\psi^{2}(n_{0}+1)^{2}}{U_{0}} \nonumber \\
&& \nonumber \\
& - &
\frac{1}{2\beta}\sum_{r=1}^2\log\left[2\cosh\left(\beta(\lambda_{r}+
\frac{\alpha_{r}}{U_{0}})\right) \right] \; ,
\end{eqnarray}
where
\begin{eqnarray}
\label{coefficients}
\lambda_{r} & = & \sqrt{(\delta-K\chi_{r})^{2}+ (J^{F,A}\phi)^{2}(n_{0}+1)} \; ,\nonumber\\
&& \nonumber \\
\alpha_{r} & = & -\frac{1}{\lambda_{r}} \bigg\{ 2 (J^{F,A}
\phi)^{2}K\psi_{r}(n_{0}+1)^{2} + \big[ (J^{F,A} \phi)^2-K^{2}
 \psi_{r}^{2}(n_{0}+1) \big] (\delta-K\chi_{r}) \bigg\} \; ,
\end{eqnarray}
\end{widetext}
In \eq{FreeEnergy} the index $r$ runs over the two sub-lattices
$S_1$ and $S_2$ in which the original lattice is split. In the
ferromagnetic case $2 |\delta|>K$ the two sublattices coincide and
$\lambda_1 = \lambda_2 = \lambda$, $\alpha_1 = \alpha_2 = \alpha$,
where
\begin{eqnarray}
\lambda & = & \sqrt{(\delta-K\chi)^{2}+
(J^{F}\phi)^{2}(n_{0}+1)}\\
\alpha & = & -\frac{1}{\lambda}\left\{ 2
(J^{F}\phi)^{2} K \psi(n_{0}+1)^{2}+ \right. \nonumber \\
& & \left.((J^{F}\phi)^{2}-K^{2}
 \psi^{2}(n_{0}+1))(\delta-K\chi)\right\} \; .
\end{eqnarray}
The free energy per site so obtained depends, obviously, on the
three order parameters $\phi$, $\psi$ and $\chi$, that must be
evaluated self-consistently. Regarding $\phi$ and $\psi$, this is
an easy task; it is accomplished by simply determining the minimum
of the free energy in each case. The existence of the minimum is
assured by the right concavity of the free energy and hence it is
enough to impose $\partial f/\partial \phi =0$ and $\partial
f/\partial \psi =0$, in order to determine their extremal values.
On the contrary, the order parameter $\chi$, that depends both on
$\phi$ and $\psi$: $\chi = \chi ( \phi, \psi )$, cannot be simply
evaluated by fixing the extremality conditions. One must instead
resort to its definition, and solve analytically for it, i.e., we
must use the fact that $\chi$ is defined as the average value of
$\hat{m}_i$ and exploit this definition to determine it. This
evaluation may be performed using the same mathematical technique
employed for the evaluation of the free energy (see the Appendix).
The self-consistent equation so obtained together with the
extremal condition with respect to the hopping order parameter are
the set of relations that are needed to analyze the phase diagram
of the system. As we have already mentioned, the expressions
derived for the Hamiltonians and the free energies are obtained
and are valid in the moderately strong coupling regime, where only
the first four lowest lying states are considered (see the
Appendix for more details). This condition is consistently met
when the ratio $w$ of the dressed hopping to the on site
interaction coupling strength does not exceed unity: $w \equiv
J(n_0 +1)/U_0 < 1$.

\section{Results and Discussion}

The study of the different possible solutions of the three
equation needed to determine the different order parameters
supplies the information needed about the phase diagram of the
system. Obviously, it is not possible to follow analytically all
the solutions as functions both of the temperature and of the
Hamiltonian parameters, and exact numerical solutions will be used
to track the phase diagram in the whole range of physical
parameters.

\subsection{Phase diagram: qualitative aspects, absence of pair 
superfluidity, and the role of many-body interactions}

As it is well known, the standard BH model sustains a phase
transition between a single-boson superfluid phase and a normal
(disordered) phase that at vanishing temperature reduces to a Mott
insulator phase \cite{Fisher,van Oosten,Sheshadri}. The first
issue we wish to address here is whether, due to the presence of
energy terms corresponding to the hopping of pairs of atoms
between adjacent sites, the EBH model can sustain a new superfluid
phase characterized by a non vanishing value of the
pair-superfluidity order parameter $\psi$ either in the absence
($\phi=0$) and/or in the presence ($\phi \neq 0$) of the standard
superfluidity of individual atoms. In fact, we find that within
the EBH setting this is never the case, and it is always $\psi=0$.
This negative result can be easily understood by looking at
the condition of extremality obtained by differentiating 
\eq{FreeEnergy} with respect to the pair-superfluidity order 
parameter for each sub-lattice. By recalling that within the
two sub-lattices ($r=1,2$), $\psi_1=-\psi_2=\psi$, one finds:
\begin{equation}
\label{derivata} 
\psi - \frac{K(n_{0}+1)^{2}}{U_{0}} \psi =
\frac{1}{4 K U_{0}}\sum_{r=1}^{2}\frac{\partial
\alpha_r}{\partial\psi} 
\tanh \left( \beta ( \lambda_r + \frac{\alpha_r}{U_{0}} ) \right) .
\end{equation}
We immediately observe that the left-hand side of \eq{derivata}
involves both terms proportional to the zero-order power of $U_0$
and to the inverse of $U_0$, while in the right-hand side only the
term proportional to the inverse of $U_0$ appears. Since our
analysis is carried out in the strong-coupling limit (large $U_0$)
and the hyperbolic tangent takes values in the range $[-1,1]$, the
extremality condition \eq{derivata} is effectively of the form
$(1-A) \psi = B \psi$. This relation, regarded as a linear
algebraic equation for the pair-superfluidity order parameter $\psi$, 
is satisfied only if $\psi$ vanishes, due to the fact that the
real coefficients $A$ and $B$ are both very small: $A,B \ll 1$
(In fact, in most situations it is even $A,B \ll 0.1$).
To illustrate how this takes place, we focus on the case in which the
single particle order parameter $\phi$ vanishes (it is easy to check
numerically that the same conclusions hold true for $\phi \neq 0$ as
well). For null single particle superfluidity, $\phi=0$,
by keeping in mind that $K=2U_0\varepsilon^2$, one has
$A = 2\varepsilon^2 (n_0+1)^{2}$ and 
$B = \varepsilon^2(n_0+1)\tanh(\beta(\lambda 
+ \frac{\alpha}{U_0}))|_{\phi=0}$. 
Let us fix, for instance,  $n_0 = 9$, and evaluate the
coefficients $A$ and $B$ at two different values of
the lattice attenuation parameter, $\varepsilon=0.01$ and
$\varepsilon=0.001$. Then, in the first case we have $A = 0.02$ 
and $B = 0.001 \tanh(\beta(\lambda+\frac{\alpha}{U_0}))|_{\phi=0} $. 
In the second case we have $A = 0.0002$ and 
$B = 0.00001\,\tanh(\beta(\lambda+\frac{\alpha}{U_0}))|_{\phi=0}$.

The circumstance according to which the pair-superfluidity order
parameter vanishes holds in general. In fact, one can show that
taking into account corrections proportional to any power $l>2$ of
$\varepsilon$, leads to new hopping terms that are
different from the single-particle and pair hoppings that we have
considered so far. An example of such hopping terms of higher
order is provided by the operator describing the collective
tunneling from, say, site $i$ to site $j$ of pairs consisting of
two atoms localized on nearest neighbor sites, and so on. Each of
these new hopping terms is associated to a suitable order
parameter to be determined self-consistently. For these order
parameters, the same arguments exploited for $\psi$ hold, implying
that the extremality conditions are always and only satisfied if
all the order parameters for the superfluidity of composite
particles vanish identically. This result leads to a behavior,
with respect to superfluidity, that is ruled by the single-atom
superfluid order parameter $\phi$, and is thus qualitatively
similar to that exhibited by the standard BH model at finite
temperature, under the effect of a superimposed harmonic
confinement \cite{GIMDS}.

From a fundamental physical point of view, the impossibility to
obtain a superfluid phase in which the tunneling particles are
composite bosons made up by two or more bosonic atoms stems from
the fact that the local eigenstates of lowest on-site energy are
always the two consecutive Fock states $\ket{n_0}$ and
$\ket{n_0+1}$. This always makes the hopping of any aggregate of
atoms energetically unfavorable. To engineer a superfluid phase of
composite bosons is thus necessary to overcome this limitation and
realize a situation in which the two lowest local eigenstates of
lowest on site energy are $\ket{n_0}$ and $\ket{n_0+k}$, with $k >
1$ being the dimensionality of the generic composite. To achieve
this goal is then necessary to engineer and take into account
many-body interactions comparable in magnitude to the standard
bilinear ones (two-body collisions) that are usually the only
interactions included in the description of dilute systems of
interacting bosons.

In the following, we will first consider the ferromagnetic model
to track the SF-PMI quantum phase transition as the
zero-temperature limit of the the finite-temperature transition
between the disordered and the superfluid phase. Later on, we will
consider the antiferromagnetic model to determine the SF-DWMI
phase transition, and, finally, we will analyze the full quantum
phase diagram of the EBH model at zero temperature.

\subsection{Finite- and zero-temperature transitions to superfluidity}

\begin{figure}
\centering
\includegraphics[width=8cm]{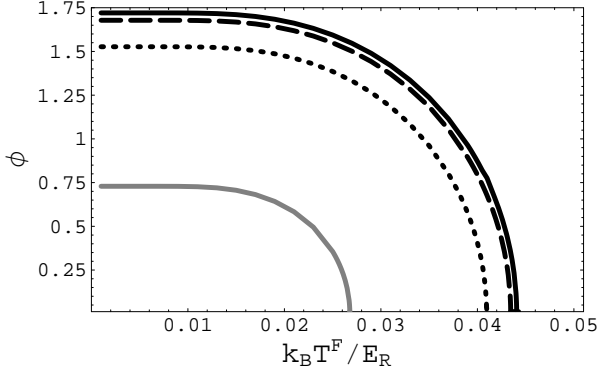}
\caption{Single atom superfluid order parameter $\phi$, as a
function of the dimensionless ``ferromagnetic'' critical
temperature $k_{B}T^{F}/E_{R}$ rescaled in units of the lattice
recoil energy $E_R$, for $J=0.01 U_{0}$. From top to bottom,
behavior for $\Delta=0$ (black solid line), $\Delta=0.009 U_{0}$
(dashed line), $\Delta=0.02 U_{0}$ (dotted line), and
$\Delta=0.04U_{0}$ (solid line). Notice, as expected, that the
critical temperature lowers as the energy gap $\Delta$ increases.}
\label{orderparameterfig}
\end{figure}

\begin{figure}
\centering
\includegraphics[width=8cm]{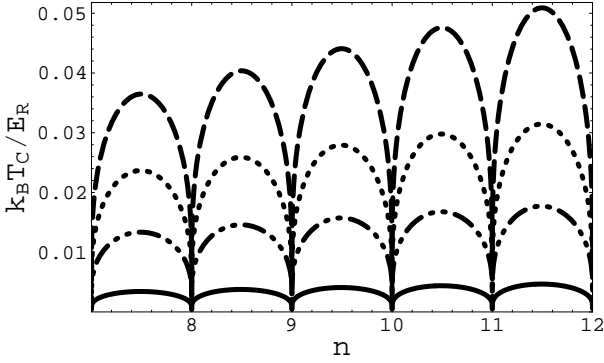}
\caption{Dimensionless critical temperature $k_{B}T_{c}^F/E_{R}$,
rescaled in units of the lattice recoil energy $E_R$, for the
transition between the disordered (high-temperature) phase and the
superfluid (low-temperature) phase as a function of the filling
factor $n$. From top to bottom, functional behavior for $J=0.01
U_{0}$ (dashed line), $J=0.007 U_{0}$ (dotted line), $J=0.004
U_{0}$ (dashed-dotted line), and $J=0.001U_{0}$ (solid line). As
the overall hopping $J$ increases, the critical temperature
rises.} \label{Tcvsn}
\end{figure}

\begin{figure}
\centering
\includegraphics[width=8cm]{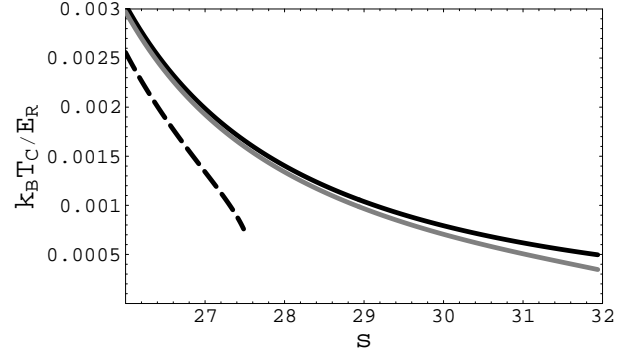}
\caption{Dimensionless critical temperature $k_{B}T_{c}^F/E_{R}$,
rescaled in units of the lattice recoil energy $E_R$, for the
transition to superfluidity as a function of the dimensionless
lattice depth $s \equiv V_0/E_R$. From top to bottom, behavior for
$\Delta=0$ (solid black line), $\Delta=0.0005U_{0}$ (solid gray
line), and $\Delta=0.0009 U_{0}$(dashed black line).}
\label{Tcvss}
\end{figure}

Starting from the ``ferromagnetic'' grand canonical free energy,
the critical diagram for single atom superfluidity is reported in
\fig{orderparameterfig} for different sets of values of the
Hamiltonian parameters.
From Fig.~\ref{orderparameterfig}, we observe a lowering of the
critical temperature $T{_{c}}^{F}$ with increasing amplitude of
the energy gap $\Delta$. From a physical point view, this
situation is due to the fact that for high values of $\Delta$,
bosons experience a high potential barrier that contrasts the
hopping from a site to its nearest neighbor with a consequently
increasing difficulty for the whole system to go toward an ordered
phase and hence the superfluid transition occurs in ``colder''
zones. Solving the equation $\partial f/ \partial\phi
|_{\psi=0}=0$ with  $\chi$ evaluated at $\psi=\phi=0$, we get
\begin{widetext}
\begin{equation}
\label{unosuTc} \beta_{{c}}^{F} \, = \, \frac{1}{\delta - H
\chi(0,0)} \tanh^{-1} \left[ \frac{2/(J-H\chi(0,0)) -
2(n_{0}+1)/U_{0}}{(n_{0}+1)/(\delta-K \chi(0,0)) - 2/U_{0}}
\right] \; ,
\end{equation}
\end{widetext}
where $H=2 U_{0} \varepsilon^{\frac{3}{2}}$. In \fig{Tcvsn} we
show the behavior of the critical temperature in units of the
lattice recoil energy $E_R$ as a function of the filling factor $n
= n_0+1/2 + \chi$ for different values of the Hamiltonian
parameters. Due to the existence of a region with density wave
order, the condition $\chi=0$ or $n=n_0+1/2$ is verified
throughout an entire (although small) region in the space of
parameters rather than at a given point in it. In this region the
critical temperature, as we will see in the following, becomes
function of the DWMI order parameter $\chi$, while the filling
factor remains constant. Hence \fig{Tcvsn} reproduces the behavior
of the critical temperature only in the ferromagnetic-like
instance, and the value of the critical temperature with
semi-integer value of the filling factor requires a longer
analysis that will be presented in the following subsection.
\fig{Tcvsn} shows the competition between thermal effects and
ordered mobility. At fixed $n$, a larger hopping amplitude
corresponds to a higher critical temperature. In \fig{Tcvss} we
show instead the behavior of the critical temperature as a
function of the optical lattice depth $s \equiv V_0/E_R$ for
different values of the chemical potential. As the depth of the
lattice increases, hopping and mobility are suppressed, and the
critical temperature of the superfluid transition lowers.

\begin{figure}
\centering
\includegraphics[width=8cm]{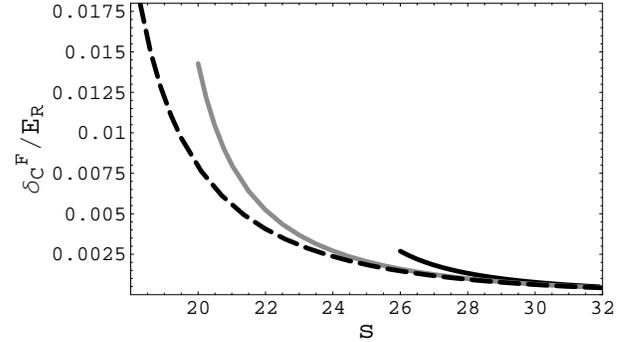}
\caption{Dimensionless critical field $\delta_{c}^{F}/E_{R}$, as a
function of the lattice depth $s$. From left to right, behavior
for $\lambda=65$ (dashed black line), $\lambda=39$ (solid gray
line), $\lambda=13$ (solid black line).} \label{Deltacvss}
\end{figure}

Looking back at \fig{Tcvsn} we must notice that, obviously, the
critical temperature vanishes for integer value of the filling
factor (no superfluidity allowed). Requiring instead that
$\beta_{c}^{F}$ assumes an infinite value in \eq{unosuTc}, we
obtain the critical condition on the local gap or ``external
magnetic field'' $\delta$ for the zero-temperature quantum phase
transition from a PMI phase to superfluidity:
\begin{equation}
\label{criticalfieldsp} \delta_{c}^{F} \, = \, \frac{(J -
H/2)(n_{0}+1)}{2\left( 1 -\frac{n_{0}(J-H/2)}{U_{0}}\right)} \, +
\, \frac{K}{2} \; .
\end{equation}
If the local \virg{gap} $\delta$ is smaller than $\delta_{c}^{F}$,
the system is in a superfluid phase; otherwise, a Mott insulator
is realized. Clearly, this result holds provided that $\delta >
\frac{K}{2}$, i.e the system cannot access the density wave
region. The behavior of $\delta_c^F$ as function of the depth of
the optical lattice is showed in \fig{Deltacvss}, for different
values of the anisotropy parameter occurring in the external
harmonic confinement.
Concerning Fig.~\ref{Tcvss}, we observe that the functional
behavior is not plotted down to small values of the lattice depth
parameter $s$. This is due to the fact that for values of $s$ in
the approximate range $[0,15]$, the weak coupling ratio $w \equiv
J(n_0+1)/U_{0}$ may exceed unity, so that the strong coupling
approximation breaks down. The graph has thus been plotted in the
interval of values of $s$ such that $0\leq w \leq 0.6$. We observe
that the critical temperature decreases for increasing $s$. This
is due to the fact that the greater the lattice depth, the more
the on-site interaction tends to dominate on the hopping. Hence,
in order to achieve the onset of the transition to the ordered
superfluid phase it is necessary to operate at lower temperatures.
Moreover, once the lattice depth is fixed, the critical
temperature is lower for larger energy gap $\Delta$, in agreement
with the results presented in Fig.~\ref{orderparameterfig}.
Concerning Fig.~\ref{Deltacvss}, each of the three curves,
corresponding to a different value of the transverse trapping
frequency (anisotropy parameter), is plotted for a different range
of the lattice depth $s$. Each of these different ranges
corresponds to to the different zones in which, for the various
values of the anisotropy parameter, the relation $0 \leq  w \leq
0.6$ holds. We see that the range of permissible values of $s$
grows with increasing anisotropy.

\subsection{Unified finite-temperature phase diagram}

The transition from an ordered superfluid phase to a Density Wave
Mott Insulator can be determined starting from the
antiferromagnetic grand canonical free energy along the same lines
followed to analyze the SF-PMI phase transition in the
ferromagnetic grand canonical setting. Hence, we will analyze the
zero-temperature SF-DWMI quantum phase transition by first
determining the ``antiferromagnetic'' critical temperature
$T_{c}^{A}$ and critical field, $\delta_{c}^{A}$, the associated
finite-temperature phase diagram, and by finally taking the
zero-temperature limit. Concerning the first step, straightforward
evaluation yields:
\begin{widetext}
\begin{equation}
\label{unosuT} 1-\frac{J\,(n_0+1)}{U_{0}} \, = \, J \sum_{r=1}^{2}
\left\{ \tanh \left[ \beta_{c}^{A} \left( \delta + K
\chi_{r}(0,0)\right) \right] \left(
\frac{n_0+1}{\delta+K\chi_{r}(0,0)}-\frac{2\big(\delta+K
\chi_{r}(0,0)\big)}{U_{0}\sqrt{\big(\delta+K\chi_{r}(0,0)\big)^{2}}}
\right) \right\} \; ,
\end{equation}
\end{widetext}
for the inverse $\beta_{c}^{A}$ of the \virg{antiferromagnetic}
critical temperature, and
\begin{widetext}
\begin{equation}
\label{cc} \delta_{c}^{A} \, = \,
\frac{J\,U_{0}(n_0+1)+\sqrt{4K^2( J\,(1+n_0)-U_{0})^2+(J\,
U_{0}(n_0+1))^{2}}}{4(U_{0}-J\,(1+n_0))}
\end{equation}
\end{widetext}
for the \virg{antiferromagnetic} critical field, that differs
analytically from the \virg{ferromagnetic} one expressed by
\eq{criticalfieldsp}. Of course, in those regions of the space of
parameters that allow to neglect the nonlocal energy terms,
$\delta_{c}^{A}$ and $\delta_{c}^{F}$ coincide exactly. The
relations expressed by \eq{unosuT} and \eq{cc} together with the
corresponding ones previously obtained for the SF-PMI phase
transition allow us to construct the full phase diagram of the EBH
model both at finite and at zero temperature.

\begin{figure}
\begin{centering}
\includegraphics[width=8cm]{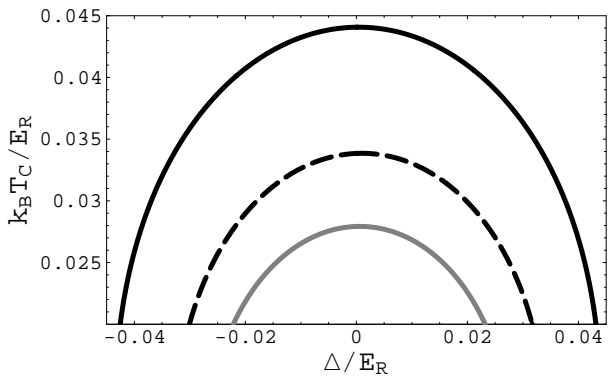}
\caption{The critical temperature $k_{B} T_{C}/E_{R}$ for the
transition from a disordered phase to the ordered superfluid phase
as a function of the local energy gap $\Delta/E_{R}$. From top to
bottom, behavior for $J=0.01 U_{0}$ (black solid line), $J=0.009
U_{0}$ (dotted line), and $J=0.007 U_{0}$ (gray solid line).}
\label{Scenario1}
\end{centering}
\end{figure}

Concerning the finite-temperature scenario, we report in
Fig.~\ref{Scenario1} the behavior of the critical temperature for
the transition from the high-temperature disordered phase to the
low-temperature ordered superfluid phase as a function of the
energy gap $\Delta$ for different values of the lattice
parameters. Recalling the relation $\delta \equiv \Delta - 2U_{0}
\epsilon^2 (n_0 + \frac{1}{2})$ that connects the local external
field with the energy gap, we can follow the entire evolution of
the critical temperature as a function of the external field,
moving smoothly through the ferromagnetic and the
antiferromagnetic regimes. The thermodynamic evolution of the
system may be considered made up of three continuous intervals.
Going from left to right on the abscissa in Fig.~\ref{Scenario1},
the interval of negative values of $\Delta$, that maps in the
interval of negative values $\delta < - K/2$, corresponds to a
ferromagnetic critical temperature $T_c = T_c^F$. The central part
of the interval, around $\Delta = 0$, corresponds to the interval
of negative and positive values $-K/2 < \delta < K/2$ and to an
antiferromagnetic critical temperature $T_c = T_c^A$. Finally, the
right part of the interval of positive values of $\Delta$ that
maps in the interval of positive values $\delta > K/2$,
corresponds again to a ferromagnetic critical temperature $T_c =
T_c^F$.

In the central interval, the finite-temperature analogue of the
zero-temperature SF-DWMI quantum phase transition is realized
(antiferromagnetic-like coupling). In the two external regions the
finite-temperature analogue of the zero-temperature SF-PMI quantum
phase transition is realized (ferromagnetic-like coupling). In the
former, antiferromagnetic-like case, the behavior of the critical
temperature as the gap varies in the range $[-K/2,K/2]$ should be
represented by a flat, constant line in the central region of
Fig.~\ref{Scenario1}, joining the two curves representing the
critical temperature in the two ferromagnetic-like external
regions. However, due to its extremely  small extension, this
\virg{antiferromagnetic connection} appears shrunk to a single
point, the overall maximum of the critical temperature. Hence, the
only visible landscape in the regions above the critical
temperature in Fig.~\ref{Scenario1} is the one relative to the
finite-temperature analogue of the zero-temperature arrangements
in which a Pure Mott Insulator arrangement is favored. In these
regions, the critical temperature lowers as the modulus of the gap
increases. This fact is in agreement with the considerations
developed for Fig.~\ref{orderparameterfig}. In particular, when
the modulus of the gap is larger than the value of the critical
field for the SF-PMI quantum phase transition, the critical
temperature vanishes and the lattice is characterized by the same,
constant integer filling factor on all sites. Moreover, by using
the same arguments presented in the previous subsection for the
behavior of the critical temperature as a function of the filling
factor, we can understand the lowering of the critical temperature
as a consequence of the lowering of the hopping amplitude from a
site to its nearest neighbor.

\subsection{Unified zero-temperature quantum phase diagram: Superfluid, Pure Mott
Insulator, and Density Wave Mott Insulator phases}

\begin{figure}
\begin{centering}
\includegraphics[width=8cm]{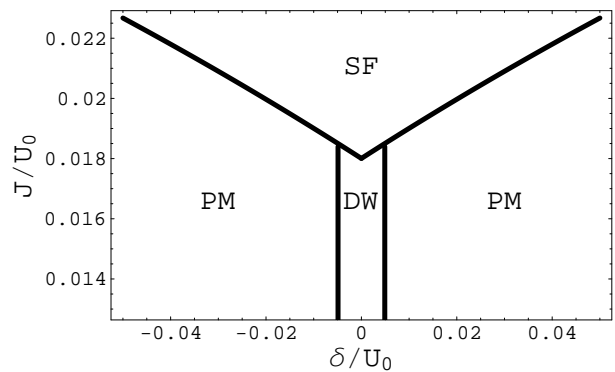}
\caption{The zero-temperature quantum phase diagram of the EBH
model in the strong coupling regime. Horizontal axis: dimensionless
gap $\delta/U_{0}$. Vertical axis: dimensionless normalized
hopping amplitude $J/U_{0}$. The vertical lines are the separation
lines between the DWMI and PMI phases. The oblique lines are the
separation lines between the SF and PMI phases and between the SF
and DWMI phases. Symmetrically placed on the sides of the cusp
point are the two tricritical points. The phase diagram is plotted
for a value of the lattice attenuation factor
$\varepsilon=0.07$ and on site occupation $n_0=9$.
Here the label PM stands for Pure Mott, DW for Density
Wave, and SF for SuperFluid.}
\label{Scenario2}
\end{centering}
\end{figure}

To conclude our study, we can now consider the full diagram of
quantum phases at zero temperature, by taking the limit $\beta_c
\rightarrow 0$. At zero temperature, the quantities that determine
the transition from a kind of ordering to another one are the
Hamiltonian parameters, that are controllable quantities. When the
ratios of these control parameters are suitably tuned, macroscopic
changes take place in the ground state of the system. These
changes give rise to the zero temperature phase diagram that we
report in Fig.~\ref{Scenario2}. Here the control parameters are
the magnitude of the nearest neighbor hopping amplitude $J$ and
the local energy gap $\delta$. In establishing the boundaries
between the different phases, we must take into account that
$\delta_{c}^{F}$ and $\delta_{c}^{A}$ essentially coincide in a
wide range of values of the lattice parameters. These two
quantities determine, respectively, the boundary lines at the
quantum phase transitions from the SF to the PMI phase, and from
the SF to the DWMI phase. The quantity $\delta_{c}$ instead
determines the boundary line at the quantum phase transition from
the DWMI to the PMI ordering. An important novelty emerges with
respect to the phase diagram of the standard BH model. In fact, in
this last case there exists only one boundary line, the one
separating the SF from the PMI phase. However, in the quantum
phase diagram of the EBH model, two further boundary lines appear.

\begin{figure}
\begin{centering}
\includegraphics[width=8cm]{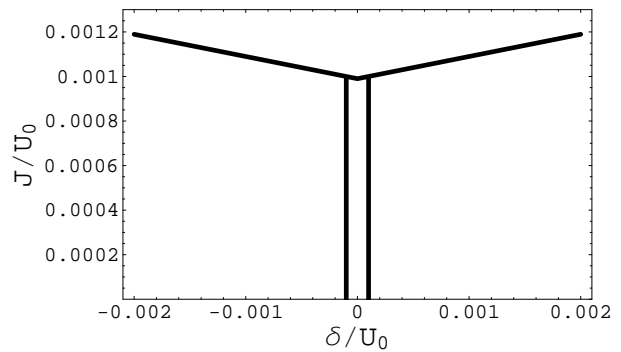}
\caption{The zero-temperature quantum phase diagram of the EBH
model in the strong-coupling regime, plotted for $\varepsilon=0.01$
and $n_0=9$. Labels denoting the various quantum
phases have been omitted, as the meaning of the oblique and vertical
lines is the same as in Fig.~\ref{Scenario2}. Notice, in particular,
the dramatic shrinking of the density wave phase for a lower value
of $\varepsilon$, compared to the one fixed in Fig.~\ref{Scenario2}.}
\label{Scenario3}
\end{centering}
\end{figure}

The first one is the coexistence curve for the SF and the DWMI
orderings; the second one is the coexistence curve for the two
insulating phases, the PMI and the DWMI. The three different
boundary lines cross at two triple points where all the three
phases coexist. From Fig.~\ref{Scenario2} and
Fig.~\ref{Scenario3}, describing the zero-temperature
phase diagram for two different values of the lattice attenuation
factor $\varepsilon$, we of course see that the zone in which the
system is in a DWMI phase is extremely small compared to the regions
occupied by the SF and PMI phases. This could be already expected
from the shrinking to a point of the corresponding antiferromagnetic-like
plateau in the finite-temperature diagram reported
in Fig.~\ref{Scenario1}.

Comparison of the two phase diagrams reported in
Fig.~\ref{Scenario2} and Fig.~\ref{Scenario3} shows that
the lattice attenuation factor plays a crucial role
concerning the area in the space of parameters in which
the system is in a DWMI phase. The smaller is the value of
$\varepsilon$ the smaller and less observable becomes the
region in which the DWMI phase takes place. This fact implies
that the experimental observation of such a phase
will require significant advances in the manipulation
and control of systems of interacting bosons in
optical lattice potentials. In particular, it will be
important to combine optical lattices potentials and
magnetic Feshbach resonances to enhance the on site
interactions to strong coupling limits, while at the
same time keeping the lattice attenuation factor in
a range of not too exceedingly small values.

\section{Conclusions and Outlook}

We have studied systems of ultracold spin-zero neutral bosonic
atoms with repulsive interactions, harmonically trapped and
regularly arranged by means of a periodic optical lattice
potential. Taking into account the series expansion of the
amplitudes of the interaction terms in powers of the lattice
potential parameters and of the lattice attenuation factor, we
have mapped the second-quantized total Hamiltonian in a new,
specific form of Extended Bose Hubbard (EBH) Hamiltonian. We have
then established various mappings of this atomic EBH model to
models of interacting spin-\mbox{$\frac{1}{2}$} systems. By using
such a correspondence, we have analyzed in a unified way the
Density Wave and Pure Mott Insulator phases supported by the
model, in analogy with the unified mean field treatment of ferro-
and anti-ferromagnetism.

We have developed the mean field theory description of the EBH
model both at finite and zero temperature, determining the free
energy density, and analyzing the finite-temperature behavior of
the model, determining the phase boundaries between the ordered
superfluid and the disordered high-temperature phase. We have
demonstrated the theoretical possibility for two different
transitions to superfluidity within the EBH model, one due to the
hopping of single atoms, and the other due to the hopping of
atomic pairs. In fact, we have given a thermodynamical proof that
only the first mechanism is realized if one truncates the
expansion in the lattice attenuation parameter at lowest order.
Finally, we have determined the zero temperature phase diagram of
the EBH model, showing the existence of a new quantum phase, the
Density Wave Mott Insulator, which is not allowed within the
framework of the standard BH model, and we have determined the
range of lattice and Hamiltonian parameters for which such a phase
can be detected. The two different forms of localized phases, Pure
Mott Insulator and Density Wave Mott Insulator, manifest
themselves, respectively, in the different behavior of the atomic
density in the lattice. The PMI phase is characterized, as usual,
by the same, constant integer filling factor throughout the entire
lattice; the DWMI is instead characterized by two different
integer filling factors in two sublattices, say $n_0$ for half of
the lattice sites, and $n_0+1$ on their neighbors (checkerboard
phase). We have studied the behavior of typical physical
quantities of the system, illustrating how the different control
parameters involved compete in determining the evolution of the
system.

Regarding future perspectives, it is to be expected that by taking
the expansion of the second-quantized total Hamiltonian further up
to higher powers in the lattice attenuation factor, a new, and
accordingly very small region in the phase diagram will emerge
where pair superfluidity, absent both in the BH and in the EBH
model, can occur, at least at extremely high filling factors, as
well as new types of intermediate range interactions and
tunneling mechanisms. The same framework introduced in this paper
may be extended to the case of systems of interacting bosons when
the excited harmonic levels of the trapping potential \cite{GIMDS}
and/or the higher-excited Bloch bands of the optical lattice
potential \cite{Scarola,Isacsson} are taken into account. It would be an
interesting challenge to further extend the scheme developed in
the present work for pure single-flavor bosons with repulsive
interactions to the case of multi-flavor bosons
and/or attractive interactions; to mixtures of bosonic and
fermionic atoms interacting on a lattice 
\cite{Faber,BoseFermi1,BoseFermi2,BoseFermi3};
and, finally, to the case of disordered and/or random optical
lattices that allow for the study of disordered ultracold atomic
gases \cite{Sanpera1,Sanpera2}.

\appendix*
\section{}

In this appendix, we will briefly review the resolvent method
\cite{Depa} needed to obtain the expression \eq{FreeEnergy} for
the free energy of the EBH model.

In order to study the thermodynamic properties of the system in
the grand canonical ensemble, we must determine the corresponding
partition function , $Z$:
\begin{equation}
\label{gp} Z \, = \, Tr[\exp(-\beta \hat{K}^{F,A}) ] \; ,
\end{equation}
where
$\hat{K}^{F,A}$ is given by \eq{grancanonicomfa} and
$\beta=1/k_{B}T$ with $k_{B}$ the Boltzmann constant. By writing
down the explicit form of $\hat{K}^{F,A}$, the grand canonical
partition function reads
\begin{eqnarray}
\label{z}
Z & \, = \,& Tr \big\{ \exp \big[ -\beta \sum_{i} \big( \hat{h}_i+ J^{F,A}\phi^{2} \nonumber \\
&& + K\psi^{2} \mp K\chi^2 \big) \big] \big\} \; .
\end{eqnarray}
In the last equation $\hat{h}_i$ represents the action of the
operator
\begin{equation}
\label{local} \hat{h} \equiv \sum_{i}\hat{h}_i \, = \,
\sum_{i}(\hat{h}_L+\hat{h}_I)
\end{equation}
on the $i$-th lattice site. As we may deduce from
\eq{grancanonicomfa}, the first operator appearing inside the sum
in the right-hand side of \eq{local} is the Hamiltonian whose
eigenstates are tensor products of local Fock states ${|n_0+k>}$
with $k$ integer or zero:
\begin{equation}
\label{L} \sum_{i}\hat{h}_L = \, \frac{U_{0}}{2}
\sum_{i}\left(\hat{m}_i^{2}-\frac{1}{4}\right) - \delta\left(
\sum_{i \in S1} {\hat{m}_i}+\sum_{i \in S2} {\hat{m}_i}\right) \, ,
\end{equation}
while
the second operator is the mean-field \virg{decoupled version} of
operators representative of the single-boson hopping, atomic-pair
hopping, and of the density-density interaction $\hat{m}_i
\hat{m}_j$, respectively:
\begin{eqnarray}
\label{I} \sum_{i}\hat{h}_I & = &- J^{F,A}
\sum_{i}\left(\hat{a}^{+}_{i}+\hat{a}_{i}\right)
\phi\nonumber\\
&& + K \chi \sum_{i \in S1} {\hat{m}_i} -(\mp K \chi) \sum_{i \in
S2}
{\hat{m}_i}\nonumber\\
&&+\frac{K}{2} \sum_{i\in S1}
\left(\hat{A}^{+}_{i}+\hat{A}_{i}\right)\psi\nonumber\\
&&-\frac{K}{2} \sum_{i\in S2}
\left(\hat{A}^{+}_{i}+\hat{A}_{i}\right)\psi \; .
\end{eqnarray}
In \eq{L} and \eq{I}, the indexes $S1$ and $S2$ denotes the two
sub-lattices in which the whole lattice is split. The meaning
of the parameters $J^{F,A}$ and $K$ have been already explained in
Sections
II and III.\\
From the grand canonical partition function, the expression for
the free energy $F^{F,A}=-\frac{1}{\beta}\ln Z$ of the system is
readily deduced. This thermodynamic potential will depend, in
general, on the mean field parameters of the theory.
\\We describe our system in the complete basis of the number
states. Keeping in mind that we are analyzing the
physics of our system in the mean field approximation,
correlations between different lattice sites are neglected and
hence, the partition function $Z$ of the
systems factorizes into the product of $M$ independent
partition functions, each of these evaluated for a single site.
In the Fock states basis and in mean field approximation
framework, the grand canonical partition function then reads
\begin{eqnarray}
\label{zbis} Z & = &
\bigg[\sum_{n=0}^{\infty}<n|\exp\big(-\beta({\hat{h}_i}+ J^{F,A}\phi^{2}+K\psi^{2}\nonumber\\
&&\mp K\chi^2)\big)|n>\bigg]^M \; ,
\end{eqnarray}
where $M$ is the total number lattice sites, the sum is in
principle performed over all Fock states, and it is intended that
the thermodynamic limit must be eventually taken. However, as
already discussed in Section III, for sufficiently strong coupling
we may limit ourselves to consider the four Fock states of lowest
energy $|n_0>$, $|n_0+1>$, $|n_0-1>$, and $|n_0+2>$ (actually, in
the ultra-strong coupling regime, it is enough to consider only
the two lowest states \cite{vanOtterlo}). This choice is fully
justified as long as the weak coupling parameter $w \equiv J(n_0 +
1)/U_0$ does not approach or exceed unity. In this way we can
determine the expressions of the physical quantities of interest
in the EBH model at first order in powers of $1/U_{0}$. To
evaluate the free energy, one needs to compute the trace of the
operator $\exp(-\beta\hat{h_i})$. However, rather than
diagonalizing $\hat{h}_i$ in the space spanned by the four
lowest-lying Fock states, it is more convenient to write the free
energy per site $f^{F,A}$ in the following way:
\begin{widetext}
\begin{eqnarray}
\label{fe1} f^{F,A} \equiv \frac{F^{F,A}}{M} & = &
J^{F,A}\phi^{2}+K\psi^{2} \mp K\chi^2
-\frac{1}{2\beta}\sum_{r=1,2}\bigg(\ln \big(\sum_{j=-1}^2
<n_0+j|\exp(-\beta \hat{h}_{i})|n_0+j>\big)\bigg)_{r} \nonumber\\
& = & J^{F,A}\phi^{2}+K\psi^{2} \mp K\chi^2
-\frac{1}{2\beta}\sum_{r=1,2} \bigg(\ln\big(\sum_{j=-1}^2
I_j\big)\bigg)_{r} \; ,
\end{eqnarray}
\end{widetext}
where
\begin{eqnarray}
\label{integrals} I_j=\frac{1}{2\pi i}\oint dz \exp(-\beta z)
G_{j,j}(z),
\end{eqnarray}
and for any couple of integers or zeroes $(j,k)$,
\begin{eqnarray}
\label{greenfunction}
G_{j,k}(z)=<n_0+j|(z-\hat{h}_{i})^{-1}|n_0+k>
\end{eqnarray}
is the Green function connecting the eigenstates
$|n_0+j>$ and $|n_0+k>$ of $\hat{h}_L$. The index $r$ appearing in
\eq{fe1} is the sub-lattice index. The operator
$(z-\hat{h}_{i})^{-1}$ appearing in the right-hand side of
\eq{greenfunction} is the so-called \virg{resolvent operator}. To
evaluate the Green functions $G_{j,k}(z)$, we have to know how the
resolvent operator acts on the ket $|n_0+k>$. According to
\eq{local}, the action of the Hamiltonian operator $\hat{h}_i$ is
nothing but the action of the operator $\hat{h}_L$ plus the action
of the operator $\hat{h}_I$. Their action can be determined
explicitly as follows. If $\hat{A}$ and $\hat{B}$ are two
operators, the following identity holds:
\begin{eqnarray}
\label{oid}
\frac{1}{\hat{A}}-\frac{1}{\hat{B}}=\frac{1}{\hat{A}}(\hat{B}-\hat{A})\frac{1}{\hat{B}}=
\frac{1}{\hat{B}}(\hat{B}-\hat{A})\frac{1}{\hat{A}} \; .
\end{eqnarray}
Then, with the identifications $\hat{A} = z - \hat{h}_i$ and
$\hat{B} = z- \hat{h}_L$, one has
\begin{widetext}
\begin{equation}
\label{oid2} \frac{1}{z-\hat{h}_i} \, = \,
\frac{1}{z-\hat{h}_L}+\frac{1}{z-\hat{h}_i}\hat{h}_I\frac{1}{z-\hat{h}_L}
\, = \, \frac{1}{z-\hat{h}_L}+\frac{1}{z-\hat{h}_L}\hat{h}_I
\frac{1}{z-\hat{h}_i} \; .
\end{equation}
\end{widetext}
The right-hand side of \eq{oid2} will be useful in the evaluation
of the propagators $G_{j,k} (z)$ in the basis formed by the
eigenstates of $\hat{h}_L$. In principle, for each values of $j$,
one needs to construct a $p \times p$ system of equations in the
variables $G_{j,k} (z)$, where the order $p$ of each system is
equal to the cardinality of the chosen basis. On the other hand,
only those functions $G_{j,k} (z)$ connecting basis vectors will
give non-vanishing contributions. Hence, in our case we have to
deal with four systems, each of these made up of four equations.
We will write down the explicit form of such a system for the
variables $G_{j,j}(z)$ that, as we can see from the \eq{fe1}, are
the needed quantities to determine the free energy per site. In
each sub-lattice labeled by $r$, and omitting the index for the
function $G_{j,j}(z)$ , we have
\begin{widetext}
\begin{eqnarray}
\label{systemfour} \delta_{j,0} & = & (z+(\delta-K\,
\chi_{r}))\,G_{0,0}(z)+\frac{J^{F,A}\,\sqrt{n_0+1}}{2}\,
\,G_{1,1}(z)\,\phi+\frac{J^{F,A}\,\sqrt{n_0}}{2}\,G_{-1,-1}(z)\,\phi \nonumber\\
&& - \frac{K \,\sqrt{(n_0+1)(n_0+2)}}{2}
\,G_{2,2}(z)\,\psi_{r} \; ; \nonumber\\
&& \nonumber\\
\delta_{j,1} & = & (z-(\delta-K \chi_{r}))\,G_{1,1}(z)+
\frac{J^{F,A}\,\sqrt{n_0+1}}{2}\,
\,G_{0,0}(z)\,\phi \nonumber\\
&& + \frac{J^{F,A}\,\sqrt{n_0+2}}{2}\,G_{2,2}(z)\,\phi-\frac{K \,
\sqrt{n_0(n_0+1)}}{2}
\,G_{-1,-1}(z)\,\psi_{r} \; ; \nonumber\\
&& \nonumber\\
\delta_{j,-1} & = & (z-U_0+3(\delta-K\,\chi_{r}))\,G_{-1,-1}(z) \nonumber\\
&& + \frac{J^{F,A}\,\sqrt{n_0}}{2}\, G_{0,0}(z)\,\phi-\frac{K \,
\sqrt{n_0
(n_0+1)}}{2} \,G_{1,1}(z)\,\psi_{r} \; ; \nonumber\\
&& \nonumber\\
\delta_{j,2} & = & (z-U_0-3(\delta-K\,\chi_{r}))\,G_{2,2}(z)+
\frac{J^{F,A}\,\sqrt{n_0+2}}{2}
\,G_{1,1}(z)\,\phi \nonumber\\
&& - \frac{K\,\sqrt{(n_0+1) (n_0+2)}}{2}\,G_{0,0}(z)\,\psi_{r} \;
\end{eqnarray}
\end{widetext}
Therefore, when $j=0$, $j=1$, $j=-1$, and $j=2$ solving system
\eq{systemfour} provides, respectively, $G_{0,0}(z)$,
$G_{1,1}(z)$, $G_{-1,-1}(z)$, and $G_{2,2}(z)$ in each
sub-lattice. Each of these solutions may be written as
\begin{equation}
\label{ND} G_{j,j} (z) \, = \, \frac{N_{j}(z)}{D(z)} \; .
\end{equation}
Since we are operating in the strong-coupling regime, we retain
only the contributions proportional to non-vanishing powers of
$U_0$ in $N_{j}(z)$ and $D(z)$. We can now obtain the explicit
expression for the free energy by direct evaluation of the second
line of \eq{fe1}. First, the integrals $I_{j}$ appearing in
\eq{integrals} are solved by the usual integration techniques in
the complex plane, and we determine the poles of the functions
$G_{j,j} (z)$ by solving the equation
\begin{eqnarray}
\label{DE} D(z) \, = \, 0 \; .
\end{eqnarray}
The roots of \eq{DE} provide the eigenvalues of the Hamiltonian
$\hat{h_i}$ at the needed order of approximation and allows to
calculate explicitly the right-hand side of \eq{integrals}.
Finally, inserting the expressions for the quantities $I_{j}$ in
the second line of \eq{fe1}, yields the desired expression for the
free energy per site \eq{FreeEnergy}.

The method of the resolvent allows as well to obtain the explicit
expression for the mean field order parameter $\chi$ in each
sub-lattice. Following the same procedure adopted to evaluate the
free energy per site, the \virg{mean number} $\chi$ in a given
sub-lattice can be determined by the the formula
\begin{widetext}
\begin{eqnarray}
\label{meanNumber} \chi & = &\frac{1}{2}
\bigg[\frac{\sum_{n}<n|\hat{n} \exp(-\beta
\hat{h_{i}})|n>}{\sum_{n}<n|\exp(-\beta
\hat{h_{i}})|n>}-(n_{0}+\frac{1}{2})\bigg] \nonumber\\
&& \nonumber \\
& = &\frac{1}{2}\bigg[\frac{n_0\,
I_{0}+(n_0+1)\,I_{1}+(n_{0}-1)\,I_{-1}+(n_0+2)\,I_{2}}{
I_{0}+I_{1}+I_{-1}+I_{2}}- (n_{0}+\frac{1}{2})\bigg] \; ,
\end{eqnarray}
\end{widetext}
where the integer index $n$ runs over the finite number of local
eigenstates being considered. The magnetization of the generic
sub-lattice $r$ finally reads
\begin{widetext}
\begin{eqnarray}
\label{meanNumberbis} \chi_{r} & = & \frac{1}{2}\bigg\{
-\frac{\delta-K
\chi_{r}}{2\lambda_{r}}\,\tanh\big[\beta\big(\lambda_{r} +
\frac{\alpha_{r}}{U_{0}}\big)\big] - \frac{(J^{F,A}\phi)^{2}
(n_{0}+1)}{2 U_{0}\lambda_{r}^{3}}\bigg[\bigg(2 K \,
(n_{0}+1)(\delta-K
\chi_{r})\,\phi^{2}\,\psi+K^{2}\,\psi^{2}\,(n_{0}+1) \nonumber \\
&& \nonumber \\
&& - (J^{F,A}\phi)^{2}\bigg)\,\tanh\big[\beta\big(\lambda_{r} +
\frac{\alpha_{r}}{U_{0}} \big)\big]\bigg]\bigg\} \; ,
\end{eqnarray}
\end{widetext}
where
\begin{widetext}
\begin{eqnarray}
\label{alpha}
\lambda_{r} & = & \sqrt{(\delta-K\chi_{r})^{2}+ (J^{F,A}\phi)^{2}(n_{0}+1)} \; \, , \nonumber\\
&& \nonumber \\
\alpha_{r} & = & -\frac{1}{\lambda_{r}}\bigg[ 2 (J^{F,A}
\phi)^{2}K\psi_{r}(n_{0}+1)^{2}+ ((J^{F,A}\phi)^{2}-K^{2}
\psi_{r}^{2}(n_{0}+1))(\delta-K \chi_{r}) \bigg] \; .
\end{eqnarray}
\end{widetext}
In the ferromagnetic-like case, the single atom hopping amplitude
is $J^{F}$, so that $\chi_{1}=\chi_{2}=\chi$, and the two
sub-lattices are characterized by the same magnetization, that is,
by the same constant filling factor $n_0$. In the
antiferromagnetic-like case, the single atom hopping amplitude is
$J^{A}$, so that $\chi_{1}=-\chi_{2}=\chi$, and the two
sub-lattices have opposite magnetizations, that is, two different
constant filling factors $n_0$ and $n_0 +1$, respectively.

\end{document}